\documentclass{article}
\usepackage{spconf,amsmath,graphicx}
\usepackage{mathrsfs}
\usepackage{amsfonts,amssymb}

\title{Efficient Cavity Searching for Gene Network of Influenza A Virus}
\name{Junjie Li, Jietong Zhao, Yanqing Su, Jiahao Shen, Yaohua Liu, Xinyue Fan, Zheng Kou$^{\dagger}$\thanks{$^{\dagger}$Corresponding authour. }}
\address{ Institute of Computing Science and Technology, Guangzhou University
}

\begin{document}
%
\maketitle
\begin{abstract}
High order structures (cavities and cliques) of the gene network of influenza A virus reveal tight associations among viruses during evolution and are key signals that indicate viral cross-species infection and cause pandemics. 
As indicators for sensing the dynamic changes of viral genes, these higher order structures have been the focus of attention in the field of virology. 
However, the size of the viral gene network is usually huge, and searching these structures in the networks introduces unacceptable delay. 
To mitigate this issue, in this paper, we propose a simple-yet-effective model named HyperSearch based on deep learning to search cavities in a computable complex network for influenza virus genetics. 
Extensive experiments conducted on a public influenza virus dataset demonstrate the effectiveness of HyperSearch over other advanced deep-learning methods without any elaborated model crafting. 
Moreover, HyperSearch can finish the search works in minutes while 0-1 programming takes days. 
Since the proposed method is simple and easy to be transferred to other complex networks, HyperSearch has the potential to  facilitate the monitoring of dynamic changes in viral genes and help humans keep up with the pace of virus mutations.

\end{abstract}
\section{Introduction}
\begin{figure}
\centering
  \includegraphics[width=0.3\textwidth]{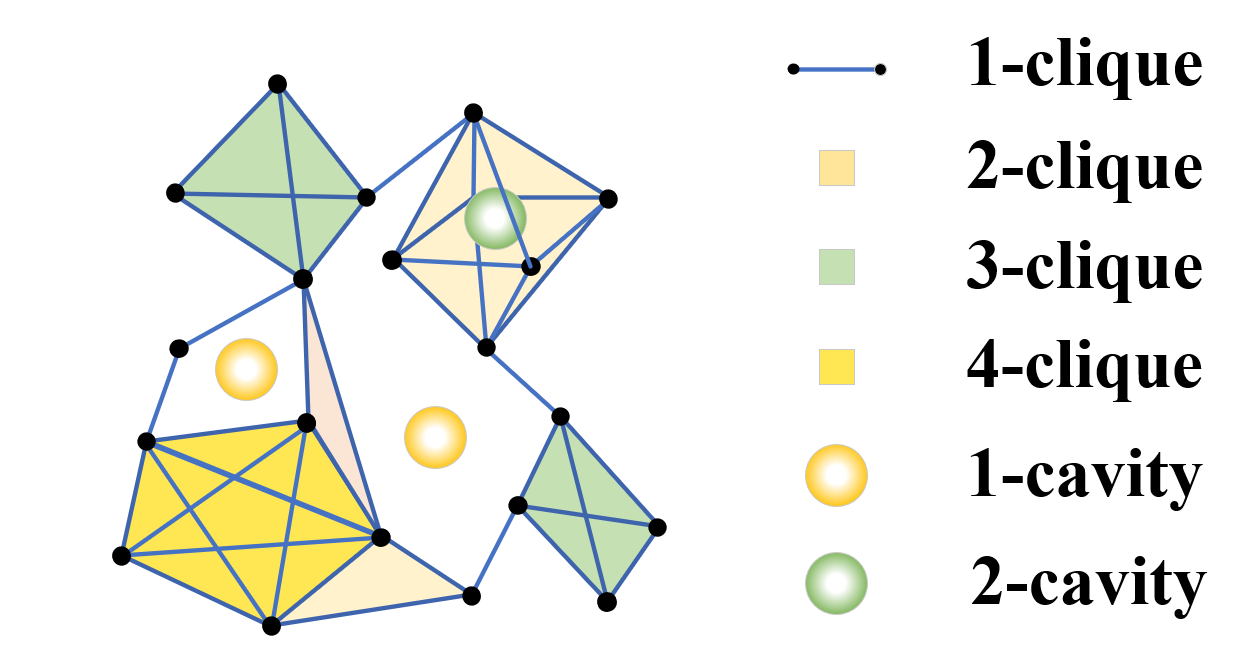}
\caption{A complex network with some cliques and high order cavities.}
\label{Fig.1}
\end{figure}

High order structures as the cliques and the cavities of complex networks are shown in Fig.\ref{Fig.1}, which often have interpretable meanings in reality. The problem of searching for higher order structures in complex networks belongs to the topological field  \cite{shi6zomorodian2004computing}\cite{shi7gu2008computational}. For instance, the cavity in complex networks maps to homology groups in algebraic topology. With the increase of computational ability these days, the significance in the reality of high order complex networks has attracted more and more attention of researchers and the theory is supplemented \cite{bookbianconi2021higher}. There are various findings relevant to the high order structure in complex networks. By exploring the brain network, humans can better understand how neurons in the brain work. By exploring animals' genetic regulatory networks (GRNs), Sinha S et al. \cite{sinha2020behavior} offered new directions for human beings to understand better of animal behaviors. By exploring plants' GRNs, De Clercq I et al. \cite{de2021integrative} predicted a series of new transcription factors (TFs) functions in Arabidopsis thaliana cells.

The other vital issue for GRNs research is how to transfer genomes (text) to GRNs (graph). X Zhang et al. \cite{zhang2012inferring} offer a solution to infer GRNs from genomes by employing the path consistency algorithm (PCA) based on conditional mutual information (CMI). Another answer came from Huynh-Thu VA et al. \cite{huynh2010inferring}, they developed a program named GENIE3, a procedure to recover GRNs from multifactorial expression data based on a variable selection with ensembles of regression trees. It can potentially capture high order conditional dependencies between expression patterns, remains acceptable computing resource consumption, and is easy to implement. 

The genome of the influenza A virus is composed of eight segments of single-strand negative RNA. With the classification of HA and NA, the influenza A virus has 16 subtypes HA and nine subtypes NA \cite{kou2021prediction}\cite{kou2022using}\cite{he2022social}. Following fast mutation of the viral genome, antigen escaping, drug-resistance, and virulence enhancing will occur \cite{kou2021prediction}\cite{kou2022using}\cite{he2022social}. In order to understand the evolution of the influenza virus, scientists tried to search high order structures by using 0-1 programming. But we are sure those approaches cannot well-solving the problem when facing a large-size network within acceptable time-consuming. Consequently, there is an urgent to find a shortcut that provides a good balance between efficiency and precision to get high order structures in complex networks.

In this work, we provide a method to find potential cavities in the gene network of the influenza A virus named Flu-Network based on HGN\cite{kdd52zhang2019hyper}. The proposed method is validated on a public influenza virus dataset which consists of 4538 viruses each containing eight segments of single-strand negative RNA. Compared with baseline (Fully-connected Neural Networks \cite{zhang2017learnability}, Convolutional Neural Networks \cite{huang2020dianet}, Graph Convolution Neural Networks, and Attention Convolutional Neural Networks \cite{huang2022lottery}), our method significantly reduced the computation time and achieved a precision of about 80\%. Importantly, we provide an algorithm that balanced computation time and precisions, so that we can further explore the biological significance represented by the cavities.

To sum up, we make the following contributions: 

(1)	A method is proposed to predict cavities in computable networks with features.

(2)	A k-clique reconstruction algorithm is proposed to filter low order features to exclude structures that obviously cannot form cavities. 


(3)	Experimental results demonstrate that the proposed method can well-balance efficiency and accuracy, achieving ~80\% precision in minutes while ordinary algorithms should take a few days.

\section{Method}
\subsection{Data processing}
\label{2.1}
The data resource in this paper is NCBI Influenza Virus Database. 
To obtain experiment data, we need to transfer genomes sequence data to GRNs which is computable. 
We build a data set named Flu-Network which consists of 4538 viruses each containing eight segments of single-strand negative RNA. The data processing process is divided into four steps:

%

\begin{figure}
\centering
  \includegraphics[width=0.5\textwidth]{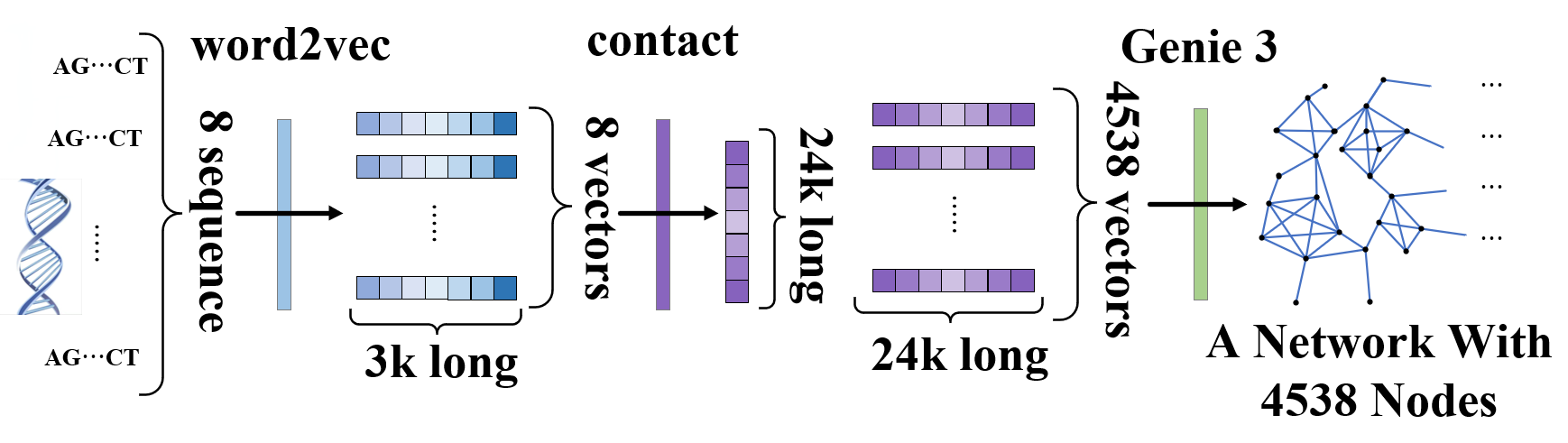}
\caption{An illustration of the word2vec program and the process of the complex network Flu-Network construction.}
\label{Fig.3}
\end{figure}


1.  First, we need to convert RNA segments to Vector by using the word2vec encoding function. Each virus data would be presented as eight vectors whose size is 1*3000. The word2vec program is illustrated in Fig.\ref{Fig.3}.

2.  Using the GENIE3 gene regulation network generation algorithm generates a complex network from the data obtained in the previous step. The process of the complex network Flu-Network construction is shown in Fig.\ref{Fig.3}.

3.  Clique and cavity search algorithms are used to search for cliques and cavities in a complex network. The searched cliques are labeled as hyper-edges, and the nodes of the searched cavities are labeled as positive samples.

Finally, after the above processing, the data of our experiment is formed.

\textbf{Searching for cliques.}
We use the common-neighbors scheme provided by D. Shi et al. \cite{dinghua2021computing} to search all cliques in the network. We define hypergraph as follows:$ \mathscr{H}_{0} = \{v^0, e^0|v^0 \in V, e^0 \in E\}$. 
where $ v^0 $ for nodes (for the network we concern, $ v^0 $ stands for a virus strands). Thus, $ V $ is the set of all the nodes.
k-order-hyperedge $ e^{0}_{k} $ is defined as follows:$ e^{0}_{k} = (v^{0}_{1},v^{0}_{2},\ldots,v^{0}_{k})$. 
where k stands for order, k-order hyperedge means the hyperlink consists of k-nodes (k-clique can be defined similarly). From now on, $ e^{0}_{k} $ stands for k-clique in the network and $ E $ contains all hyperedges.

\textbf{Searching for cavities.}
We now establish boundary matrix $B_{k}$ by cliques. Then we apply 0–1 programming to $ B_{k} $ to search all the cavities in the networks. We assign a positive label to the node which participates in the formation of cavities and otherwise a negative label.
Then the dataset is prepared and ready for the next step.

\subsection{Model}
The problem of searching for cavities in the network is an NP-complete problem. Currently, the general algorithm is to find the cavity structure by 0-1 programming, but this algorithm requires a large computational overhead. To this end, we propose a deep learning method based on representation learning, which can find cavity structures in complex networks faster in terms of efficiency on the basis of sacrificing a certain precision. The cross-entropy loss function is used to calculate the error. This method significantly reduces the computational time and still achieves a precision of ~80\%, the performance would be described in the experiment part. 

According to the definition \cite{bookbianconi2021higher}, each k-cavity is constructed by k-cliques, we only need to find all k-cliques that form k-cavities. Since the boundary matrix $ B_{k} $ can reflect the $\text{k-1}$ order connections between k-cliques forming the k-cavities\cite{dinghua2021computing}. For k-cliques in the same k-cavity, if and only if two adjacent k-cliques exhibit a connection of order $\text{k-1}$ between them in the boundary matrix $ B_{k} $. Therefore, we only need to obtain the $\text{k-1}$ order connections between these k-cliques that can form k-cavities through the boundary matrix $ B_{k} $ \cite{shi42013searching}\cite{shi5shitotally}, and then all k-cavities composed of these k-cliques can be found. Based on this idea, we use a deep learning model to learn the high order representation of the k-hypergraph reconstructed by the k-clique to distinguish whether the k-clique is a component of the k-cavities or not. So that we find all k-cliques that can form each k-cavities.

Based on the above idea, we create a method that consists of three parts: k-clique-reconstruction, hypergraph representation learning, and prediction outcomes.

\begin{figure*}
  \includegraphics[width=1\textwidth]{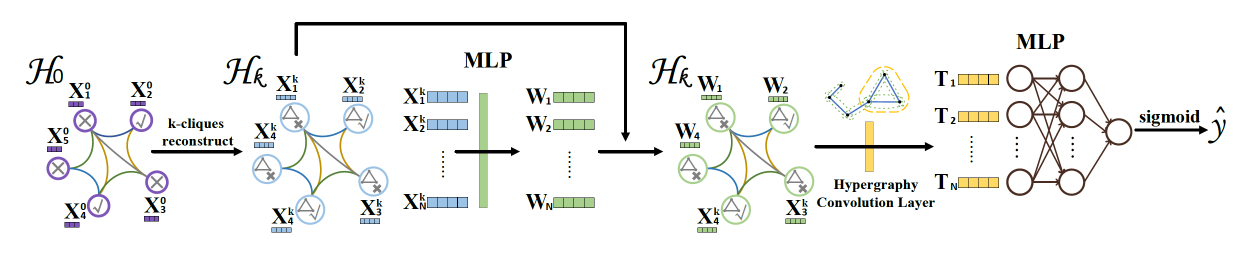}
\caption{The flow chart of the HyperSearch method. It includes 2 principal modules: a k-clique-reconstruction module and a hypergraph convolution module.}
\label{Fig.4}
\end{figure*}

\textbf{k-clique-reconstruction}
To filter out low-order features, we perform "k-clique-reconstruction" on the hypergraph $ \mathscr{H}_{k} $ obtained in Section \ref{2.1} to generate a new hypergraph: k-hypergraph. 
The generation process of the k-hypergraph is shown in Fig.\ref{Fig.4}.

Firstly, we define the expression as follows:
(a)	If node $ v_{i} $ is contained in hyperedge $ e^{0}_{k} $, then we mark it as $ v_{i} \in e^{0}_{k} $.
(b)	If all nodes in hyperedge $ e_{k_{i}} $ are contained in hyperedge $ e_{(k+p)_{j}} $, then we mark it as $ e_{k_{i}}\subset e_{(k+p)_{j}} $.

Now we define the k-clique-reconstruct algorithm$ \Phi_k $ and $ k-$hypergraph $ \mathscr{H}_{k}$ as follow: 
$  \mathscr{H}_{k} = \Phi_{k}(\mathscr{H}_{0}) = \{v^{k},e^{k}|v^{k} \in V_{k},e^{k} \in E_{k}\}$.
Where $ v^{k}_{i} = e^{0}_{k_{i}},i = 1,2,\ldots,m_{k} $. $m_{k} $ represents the number of k-clique in the hypergraph $ \mathscr{H}_0 $. As the k-cliques in the network,the node $ v^{k}_{i}$ in k-hypergraph $ \mathscr{H}_k $ are also equal to hyperedge $e^{0}_{k_{i}}$ in a hypergraph $ \mathscr{H}_0 $. As for the hyperedge $ e^{k} = e^{k}_{p} = (v^{k}_{1},\ldots,v^{k}_{p})$, ${p}$represents ${p}$-order hyperegdge which contains ${p}$ nodes. 
Where $ v^{k} \in e^{k}_{p} $, there exists $ e^{0}_{k+p} = (v^{0}_{1},v^{0}_{2},\ldots,v^{0}_{k+p}) \in E $, such that
$ v^{k} = e^{0}_{k} \subset e^{0}_{(k+p)}$.
For each node $ v^{k} $ in the hypergraph $ \mathscr{H}_k $, there are $k$ nodes in the hypergraph $ \mathscr{H}_0 $.

As for the vector $ v^{k}_{i} = e^{0}_{k_{i}} =(v^{0}_{1},v^{0}_{2},\ldots,v^{0}_{k})$ of each node in the hypergraph $ \mathscr{H}_k $, we obtain the vector $ v^{k} $ of all nodes by concatenating the vectors $ v^{0} $of all nodes in the hypergraph $ \mathscr{H}_0 $, i.e.
$ \text{Vector}(v^{k}_{i}) = \text{concat}(\text{Vector}(v^{0}_{i}),i = 1,2,\ldots,k)$. 
After k-clique reconstruction, we get the new hypergraph $ \mathscr{H}_k $ and $ \{X^{k}_{i} = \text{Vector}(v^{k}_{i})\} $.

\textbf{Multilayer Perceptron module.}
In order to learn the features in the node vector $ \{X^{k}_{i} = \text{Vector}(v^{k}_{i})\} $ from the hypergraph. We use the MLP module to process the vector $ \{X^{k}_{i}$ and get the processed vector $ W_{i} $, i.e. $ W_{i} = \text{MLP}(X^{k}_{i}) $. Combined with the binary cross entropy loss function, we expect the resulting vector $ W_{i} $ to capture all the features of the node vector $ X^{k}_{i} $, which allows us to exclude confounding features in the node vector $ X^{k}_{i} $.

\textbf{High order features capturing module.}
The input of the module is the hypergraph $ \mathscr{H}_k $ which is reconstructed by $k$-clique and the vector $ W_{i} $ which is processed by MLP to capture high order features. Such that we can get the feature-vector of the k-clique when it is one of the boundary cliques of a k-cavity.

We train a high order feature extraction function $ \Psi(\cdot) $ to catch high order feature$ T_{i} $, i.e. $ T_{i} = \Psi(W_{i},\mathscr{H}_k) $. As shown in Fig.\ref{Fig.4}, $ \Psi(\cdot) $consists of hypergraph convolution modules \cite{kdd5bai2021hypergraph}\cite{kdd44yadati2019hypergcn} and hypergraph attention modules \cite{kdd5bai2021hypergraph}\cite{kdd11ding2020more}\cite{kdd52zhang2019hyper}. Where the hypergraph convolution operator captures higher order features and the attention operator weights these higher order features. To learn the high order structure features in the hypergraph $ \mathscr{H}_k $ for each node, we propagate high order features with a hypergraph convolutional layer. Vanilla Laplacian matrix is introduced here:$ L = D^{-\frac{1}{2}} H \boldsymbol{B}^{-1} H^{T} \boldsymbol{D}^{-\frac{1}{2}}$. 
where $ \boldsymbol{D} \in \mathbb R^{n\times n} $ is a diagonal matrix representing the degree of the nodes (The degree of the node indicates how many hyperedges attach to the node). And $ \boldsymbol{B} \in \mathbb R^{m\times m} $ is a diagonal matrix used to represent the number of nodes contained in each hyperedge $ e^{k}$ of the hypergraph $ \mathscr{H}_k $. 

The hypergraph convolutional operator is defined as 
$ \boldsymbol{T}^{l+1} = \text{ReLU}(\boldsymbol{LT}^{l}\boldsymbol{W}^{(l+1)})$. 
where $ \boldsymbol{T}^{l} $ is the feature-vector from the $l^{th}$ hypergraph convolutional layer. $ \boldsymbol{W} \in \mathbb R^{d^{l}\times d^{(l+1)}} $ is the feature-matrix of the $(l+1)^{th}$ hypergraph convolutional layer. Respectively, $d^{l}$ and $d^{(l+1)}$ denote the feature dimensions of the $d^{th}$ and $(d+1)^{th}$ layers.

Then, the hypergraph attention mechanism is used to assign weight to every hyperedge $ e^{k}_{i} $ in hypergraph $ \mathscr{H}_k $, i.e. $ \hat{e}^{k}_{i} = \alpha_{i}e^{k}_{i} $. We want to capture information about features on important hyperedges, especially, when k-clique is a component of k-cavity we expect to boost weights in the k-clique. Then we can better distinguish those k-clique. Denote feature vectors from the last hypergraph convolutional layer as follow:
$ \boldsymbol{T} = {\{\boldsymbol{T}_{i}\}}^{m_{k}}_{i=1}$.
where $ \boldsymbol{T}_{i} $ refers to the hypergraph convolutional feature vector of each k-clique in the Flu-Network.
\textbf{Prediction outcomes.}
We model the potential outcomes of the $i^{th}$ k-clique as $ {\hat{y}}^{k}_{i} = f(\boldsymbol{T}_{i})$.
Where the $f$ is a learnable function to predict the potential outcomes. $f$ is a module constructed by two-layers- MLP and a sigmoid function with the binary cross entropy loss function. E.g. $ f(\boldsymbol{T}_{i}) = \text{sigmoid}(\text{MLP}_{2}(\boldsymbol{T}_{i})) $.

\section{Experiment}
\subsection{Dataset}
We use the dataset obtained in section \ref{2.1} which contains 4538 viruses, each virus is presented by 8 vector that is 3k long, and a hypergraph $ \mathscr{H}_0 $ that describe the relationship between viruses. 
In this experiment, we only consider 1-cavities. We found 83 1-cavities in this network. A total of 277 virus samples were marked as positive samples, and the rest were marked as negative samples. 

\subsection{Evaluation Metrics}
In this study, since our goal is to find positive samples, we choose the general evaluation metrics of Accuracy and Precision. Meanwhile, the dataset we use for performance is imbalanced. Therefore, we use area under the receiver operating characteristic curve (AUROC) and the area under the precision-recall curve (AUPR) as the metrics to evaluate our model. The closer the value of AUROC is to 1, the better the performance of the model is. Similarly, the closer the value of AUPUR is to 1, the better the model effect is. 

\subsection{Comparison with baseline model}
We choose GCN, CNN, CNN with attention mechanism, and MLP as the baseline model to compare with our HyperSearch model. We divided the dataset into 30\% of the training set, 50\% of the test set, and 20\% of the validation set. Each model trains for 500 epochs. A desktop with a CPU (Intel Core-i7) and a GPU (NVIDIA GTX-2070) supported all the experiments. 
The comparison results are shown in Table.\ref{Table 1}. Compare to baseline models, the proposed HyperSearch has a better performance but the accuracy and precision are not as ideal as 0-1 programming. As we mentioned early, HyperSearch is a trade-off between efficiency and precision. Compared to methods that take days to obtain ground truth, our model achieves ~80\% accuracy in minutes. HyperSearch will help academics efficiently search for cavities in Flu-Network with high confidence. 

\begin{table}[]
\caption{ Comparison of results with the baseline model.}
\centering
\label{Table 1}
\resizebox{1\linewidth}{!}{
\begin{tabular}{lcccc}
\hline
Model         & \multicolumn{1}{l}{Accuracy} & \multicolumn{1}{l}{Precision} & \multicolumn{1}{l}{AUROC} & \multicolumn{1}{l}{AUPR} \\ \hline
HyperSearch   & \textbf{0.800}               & \textbf{0.775}                & \textbf{0.798}            & \textbf{0.738}           \\
GCN           & 0.707                        & 0.687                         & 0.705                     & 0.650                    \\
CNN           & 0.537                        & 0.375                         & 0.540                     & 0.365                    \\
MLP           & 0.623                        & 0.498                         & 0.630                     & 0.445                    \\
CNN Attention & 0.677                        & 0.726                         & 0.677                     & 0.649                    \\ \hline
\end{tabular}
}
\end{table}
We can see that the four evaluation metrics of HyperSearch are better than other baseline models. Compared to other baseline models, our model is able to find positive samples more accurately. At the same time, we note that both the AUROC value and AUPR value of HyperSearch are at a high level even when the sample dataset is imbalanced, implying that our model is able to recognize features of higher order structures well.
By comparing with the 0-1 programming k-cavity search method \cite{dinghua2021computing} which takes days to perform the k-cavities search. HyperSearch takes only a few minutes and achieves an accuracy of 0.79 and 0.77 on the same dataset. Even with a small loss of accuracy, our algorithm still achieves a significant improvement in terms of reduced time consumption.

\subsection{Ablation study}
The results of ablation studies are shown in Table\ref{Table 2} and Table\ref{Table 3}. We evaluate the impacts on the model when the following conditions changed.
(1)	Default setting of HyperSearch;
(2) HyperSearch without HGCN Attention;
(3)	Replace the HGCN module of HyperSearch with GCN;
(4)	HyperSearch without k-clique-reconstruction;
(5)	Different levels (4,6,8,10) of the second MLP module in HyperSearch.
Compared with the general GCN modules, the HGCN module is good at extracting high order features. By analyzing the results of whether the k-clique-reconstruction module is included in the model or not, we are convinced that the k-clique-reconstruction module is critical for model accuracy. With or without modules, there is only a 1\% difference in accuracy, while there is a 4\% difference in precision and a 12\% difference in AUPR. Table\ref{Table 3} shows the performance of different activation functions in the last MLP module. Although the activation function in MLP can improve the Precision of the model, other metrics of HyperSearch are still the best.

\begin{table}
\centering
\caption{ Ablation studies of each module to explore the contribution.}
\label{Table 2}
\resizebox{1\linewidth}{!}{
\begin{tabular}{lcccc}
\hline
Model                 & \multicolumn{1}{l}{Accuracy} & \multicolumn{1}{l}{Precision} & \multicolumn{1}{l}{AUROC} & \multicolumn{1}{l}{AUPR} \\ \hline
HyperSearch           & \textbf{0.800}               & \textbf{0.775}                & \textbf{0.798}            & \textbf{0.738}           \\
Wtihout Attention     & 0.788                        & 0.760                         & 0.787                     & 0.721                    \\
replace to GCN        & 0.736                        & 0.637                         & 0.721                     & 0.532                    \\
Without K-con         & 0.788                        & 0.732                         & 0.767                     & 0.615                    \\
MLP-end=4             & 0.735                        & 0.682                         & 0.731                     & 0.666                    \\
MLP-end=6             & 0.760                        & 0.736                         & 0.757                     & 0.698                    \\
MLP-end=8             & 0.768                        & 0.753                         & 0.769                     & 0.708                    \\
MLP-end=10            & 0.755                        & 0.738                         & 0.752                     & 0.697                    \\ \hline
\end{tabular}
}
\end{table}

\begin{table}[]
\centering
\caption{ Add different activation function in MLP.}
\label{Table 3}
\resizebox{1\linewidth}{!}{
\begin{tabular}{lcccc}
\hline
Model        & Accuracy       & Precision      & AUROC          & AUPR           \\ \hline
HyperSearch  & \textbf{0.800} & 0.775          & \textbf{0.798} & \textbf{0.738} \\
Add ReLU     & 0.785          & 0.781          & 0.786          & 0.725          \\
Add SoftMax  & 0.655          & 0.722          & 0.654          & 0.613          \\
Add Tanh     & 0.787          & 0.793          & 0.787          & 0.732          \\
Add Softplus & 0.757          & \textbf{0.813} & 0.760          & 0.724          \\ \hline
\end{tabular}
}
\end{table}

\section{Conclusion}
In this paper, we provide a method to discover potential cavities in Flu-Network based on hypergraph representation learning. We propose a HyperSearch algorithm based on the k-clique-reconstruction module and hypergraph convolution module. This algorithm has been verified in experiments that the algorithm can be used to search for high order cavity structures in a computationally complex network. By analyzing the ablation studies, the k-clique-reconstruction module which we provide in this paper can improve the performance of general hypergraph convolution modules.


\vfill\pagebreak


\bibliographystyle{IEEEbib}
\bibliography{refs}

\end{document}